\documentclass[aps,prl,twocolumn,superscriptaddress,showpacs,amsmath,amssymb,floatfix]{revtex4}
\usepackage{float}
\usepackage{epsfig}
\usepackage{amsmath}
\usepackage{amssymb}
\usepackage{bm}
\usepackage{graphicx}
\usepackage[pdftitle={Dense monoenergetic proton beams from chirped laser-plasma interaction},colorlinks=false, breaklinks=true]{hyperref}
\usepackage{breakurl}

\begin{document}

\author{Benjamin J. Galow}
\affiliation{Max-Planck-Institut f\"{u}r Kernphysik, Saupfercheckweg 1,
69029 Heidelberg, Germany}
\author{Yousef I. Salamin}
\affiliation{Max-Planck-Institut f\"{u}r Kernphysik, Saupfercheckweg 1,
69029 Heidelberg, Germany}
\affiliation{Department of Physics, American University of Sharjah, POB 26666, Sharjah,
United Arab Emirates}
\author{Tatyana V. Liseykina}
\affiliation{Institut f\"ur Physik, Universit\"at Rostock, 18051 Rostock, Germany}
\author{Zolt\'an Harman}
\affiliation{Max-Planck-Institut f\"{u}r Kernphysik, Saupfercheckweg 1,
69029 Heidelberg, Germany}
\affiliation{ExtreMe Matter Institute EMMI, Planckstrasse 1,
64291 Darmstadt, Germany}
\author{Christoph H. Keitel}
\affiliation{Max-Planck-Institut f\"{u}r Kernphysik, Saupfercheckweg 1,
69029 Heidelberg, Germany}

\pacs{52.38.Kd, 37.10.Vz, 42.65.-k, 52.75.Di, 52.59.Bi, 52.59.Fn, 41.75.Jv,
87.56.bd}

\title{Dense monoenergetic proton beams from chirped laser-plasma interaction
}

\begin{abstract}

Interaction of a frequency-chirped
laser pulse with single protons and a hydrogen gas target is studied
analytically and by means of particle-in-cell
simulations, respectively.
Feasibility of
generating ultra-intense ($10^7$ particles per bunch) and phase-space collimated
beams of protons (energy spread of about 1$\%$) is demonstrated. Phase
synchronization of the protons and the laser field, guaranteed by the
appropriate chirping of
the laser pulse, allows the particles to gain sufficient kinetic energy
(around $250$ MeV)
required for such applications as hadron cancer therapy, from
state-of-the-art laser systems of intensities of the order of $10^{21}$~W/cm$^2$.

\end{abstract}

\maketitle

Interaction of high-intensity lasers with solid targets
has attracted much interest over the past decade, due to its potential
utilization in laser acceleration of particles
\cite{plasma1,plasma2,plasma3,plasma4,plasma5,plasma6,plasma7,plasma8,plasma11,macki,mulser,liseykin,silva}.
This has given much needed impetus to efforts directed at replacing conventional
accelerators in the near future by compact and relatively low-cost devices based on an
all-optical acceleration mechanism \cite{dunne-pers}. In particular, hadron
cancer therapy \cite{cancer,debus1} may benefit from this trend.

Several regimes are now in existence for laser-plasma acceleration
\cite{mourou}.
For laser intensities of 10$^{18}-10^{21}$ W/cm$^2$ and solid targets with a
thickness ranging from a few to tens of micrometers, target normal sheath
acceleration (TNSA) is the prevailing mechanism. In TNSA, a strong quasi-static
electric field is induced at the rear surface of a thin foil,
as a result of emission and acceleration of the electrons caused by an intense linearly
polarized laser field. Ion acceleration, by the sheath electric field, has been
extensively studied
\cite{plasma1,plasma2,plasma3,plasma4,plasma5,plasma6,plasma7,plasma8,plasma11,macki}.
The regime of radiation-pressure-dominant acceleration (RPA), has
become recently accessible by decreasing the target thickness. Circularly
polarized lasers at normal incidence have been employed, which suppress electron heating and let all
particle species co-propagate as a quasi-neutral plasma bunch in front of the
light wave ("light sail"-mechanism) \cite{macchi}. Despite recent experimental \cite{habs}
and theoretical \cite{bulanov} progress,
clinically useful ion beams \cite{eickhoff} have not yet been produced within a
scheme which operates at currently available laser parameters.

In this Letter, we demonstrate the theoretical feasibility of creating proton
beams, of unprecedented energy and quality, from illuminating a hydrogen gas target
with an appropriately chirped laser pulse of intensity accessible by
state-of-the-art laser systems \cite{bahk}.
The basic idea of our model
stems from the realization that an incoming highly relativistic laser pulse
quickly ionizes hydrogen in the cell and accelerates the electrons away from the
much heavier protons, as the pulse intensity rises. At high
enough laser intensities the protons get accelerated directly by the laser field.
Chirping of the laser pulse ensures
optimal phase synchronization among the protons and the laser field
and leads to efficient proton energy gain from such a field.

We derive an expression
for the energy transfer during interaction of a particle with
chirped unfocused as well as pulsed electromagnetic fields. Results obtained
analytically for the particle's energy gain will be supported by further
simulations which describe the focused fields more accurately than the simple
plane-wave model.
Our two-dimensional (2D) particle-in-cell (PIC) simulations reveal that proton beams of energy
around $250$ MeV, energy spread of about $1\%$, and density
of $10^7$ particles per bunch, can be produced.
Beams of such quality may
potentially be suitable for use in hadron cancer therapy assuming the laser-plasma interaction
can take place close to the patient \cite{weichsel}. Following acceleration, the beams have to be collimated and guided \cite{Schollmeier}
to compensate larger variations in the laser pulse and its chirp which is still challenging to control at high intensities.
Future laser systems at the ELI or HiPER  facilities
\cite{eli,hiper} may be utilized to produce monoenergetic multi-GeV proton beams.

A point particle of mass $M$ and charge $Q$ acquires relativistic energy and
momentum, respectively, of ${\cal E}=\gamma Mc^2$ and $\bm{p}=\gamma
Mc\bm{\beta}$, where $\bm{\beta}$ is the velocity of the particle scaled by $c$,
the speed of light in vacuum, and $\gamma=(1-\beta^2)^{-1/2}$, when interacting
with the fields $\bm{E}$ and $\bm{B}$ of an intense laser pulse.
It suffices in many applications to represent the fields of the beam by plane
waves. A plane wave propagating along the $z-$axis and polarized along the $x-$axis, may be represented
by $\bm{E}=\hat{\bm{x}}E_0f$ and $\bm{B}=\hat{\bm{y}}E_0f/c$. The dependence
in $f$ on the space-time coordinates is through the combination $\omega t-kz$,
with $\omega$ the frequency and $k=\omega/c$ the wave number.
This leads to a break-up of the energy-momentum transfer equations (SI units)
 into four component equations, namely
\begin{equation}
    \label{px} \frac{d}{dt}(\gamma\beta_x) =a\omega_0 (1-\beta_z)f,\quad
\frac{d}{dt}(\gamma\beta_y) = 0,
\end{equation}
\begin{equation}
    \label{pz} \frac{d}{dt}(\gamma\beta_z) = a\omega_0 \beta_xf,\quad
\frac{d\gamma}{dt} = a\omega_0 \beta_xf,
\end{equation}
where $a=QE_0/(Mc\omega_0)$ is a normalized laser field strength and $\omega_0$
will be defined below. From the second of Eqs. (\ref{px}) we immediately
identify a first constant of the motion, namely $\gamma \beta_y=c_1$.
Furthermore, the constant $\gamma(1-\beta_z)=\gamma_0(1-\beta_{z0})=c_2$ may be
arrived at by subtracting the second of Eqs. (\ref{pz}) from the first, and
integrating. Substituting the constants $c_1$ and $c_2$ via $\beta_y$ and $\beta_z$ into
$\gamma^{-2}=1-(\beta_x^2+\beta_y^2+\beta_z^2)$, the following key expression
for the energy of the particle, scaled by its rest energy, may be arrived at
\begin{equation}\label{gamma1}
\gamma=\frac{1+(\gamma\beta_x)^2+c_1^2+c_2^2}{2c_2}.
\end{equation}

\begin{figure}
\includegraphics[width=8cm]{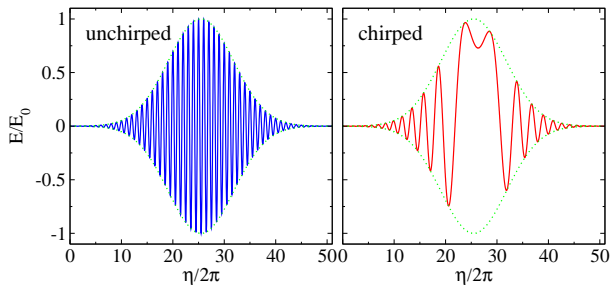}
\caption{(color online). The normalized electric field at focus of a plane-wave pulse employing
Eq. (\ref{f}). The parameters used are: $\lambda=1~\mu$m, and $\tau=50$ fs, $\phi_0=0$ and $\bar{\eta}=4s$.}
\label{fig1}
\end{figure}
The combination $\eta=\omega_0(t-z/c)$ will be used often below. Note
that with the help of $d\eta/dt=\omega_0(1-\beta_z)$,
the first of Eqs. (\ref{px}) may be formally integrated to give
\begin{equation}
\label{gbx} \gamma\beta_x=\gamma_0\beta_{x0}+a\int_{\eta_0}^\eta f(\eta')d\eta',
\end{equation}
where $\beta_{x0}=\beta_x(\eta_0)$ and $\gamma_0=(1-\beta_0^2)^{-\frac{1}{2}}$.

Frequency chirping amounts to letting $\omega$ vary with time in some fashion.
Experimentally, laser pulses with a near-uniform spectral intensity over
two octaves are feasible \cite{chirp-ex}. The phase-coherent synthesis of
separate femtosecond lasers \cite{chirp-ex-sci1} and the recent synthesis
of multiple optical harmonics \cite{chirp-ex-sci2} put forward lasers with an
even wider frequency range.
We will work with the linear chirp $\omega=\omega_0+b_{0}(t-z/c)$  \cite{chirp0,chirp1,chirp2},
with $\omega_0$ denoting the frequency at $t=0$ and $z=0$, and $b_{0}$ having a
unit of $s^{-2}$. The chirped
frequency thus becomes $\omega = \omega_0(1+b\eta)$, where
$b=b_0/\omega_0^2$ is a dimensionless chirp parameter. Also, dependence
of the fields on the space-time coordinates may be rewritten as $\omega t-kz \to
\eta+b\eta^2, $
and the chirped field function $f(\omega t-kz) \to f(\eta+b\eta^2)$. We will
initially work with
\begin{equation}\label{f}
  f = \cos(\phi_0+\eta+b\eta^2)g(\eta);\quad g(\eta)=\exp\left(-\frac{(\eta-\bar{\eta})^2}{2s^2}\right),
\end{equation}
where $\phi_0$ is a constant initial phase, $g(\eta)$ a pulse-shape function, and $s$ is related to the pulse duration $\tau$ (full-width at half-maximum) via
$s=\omega_0\tau/(2\sqrt{2\ln2})$. In our calculations, 
we choose a shift in 
$\eta$ in the
envelope function $g(\eta)$ denoted by $\bar{\eta}=4s$.
Eq. (\ref{f}) approximates an actual laser pulse in which the non-vanishing 
pulse integral may be compensated by a quasi-static tail.

To illustrate the mechanism of acceleration, we show in Fig. \ref{fig1} the
normalized electric field $E/E_0$ as a function of $\eta$.  Interaction of a
particle with the unchirped pulse will result in no energy gain, due to the plane-wave symmetry:
gain from a positive part of the field gets
canceled by loss to an equally strong negative part.
However, the chirping breaks this symmetry. Thus,
interaction with the low-frequency and strong positive part of the chirped pulse, which
extends over roughly one half of the pulse duration, results in net energy gain.

For a particle initially ($\eta=\eta_0$) at rest, $c_1=0$ and $c_2=\gamma_0=1$.
Thus, Eqs. (\ref{gamma1}) and (\ref{gbx}) give the following expression for
evolution in $\eta$ of the particle kinetic energy
\begin{equation}\label{K}
K(\eta)=(\gamma-1)Mc^2=\frac{Mc^2}{2}(\gamma\beta_x)^2.
\end{equation}
Using the initial conditions and Eqs. (\ref{gbx}) and (\ref{f}), Eq. (\ref{K}) can be analytically integrated in terms of a
complex error function. Thus, Eq. (\ref{K})
gives the kinetic energy of the particle explicitly, and helps us determine the
value of the chirp parameter $b$ that maximizes it.
Note from Eqs. (\ref{gbx}) and (\ref{K}) that the proton
energy scales linearly with the laser intensity.

\begin{figure}[t]
\includegraphics[width=8cm]{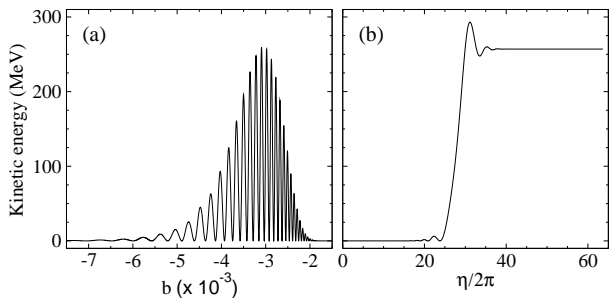}
\caption{(a) Exit kinetic energy gained by one proton as a 
function of the dimensionless chirp
parameter $b$ in the limit of large $\eta$.
(b) Evolution of the proton kinetic energy with $\eta/2\pi$ for optimal $b=-0.002979$. 
Peak power of the focused laser system
is 1 PW, which corresponds to a peak intensity of $2.55\times10^{21}$ W/cm$^2$
for $w_0=5\lambda$, $\tau=50$ fs and $\lambda=1~\mu$m. Other parameters used
are: $\phi_0=0$, and $\bar{\eta}=4s$.}
\label{fig2}
\vspace{-2mm}
\end{figure}

In order to test the applicability of the analytic plane-wave model, we compare results based on it
with those stemming from the use of focused laser fields.
For the focused fields, the Lax series expressions will be
adopted, according to which all components $E_x, E_y, E_z, B_y$, and $B_z$
\cite{sal-apb} are given in powers of the diffraction angle
$\varepsilon=\lambda/\pi w_0$, where $w_0$ is the waist radius of the beam at
focus. Simulations are performed to solve the equations of motion
numerically for a single proton submitted to the laser fields from an initial
position of rest at the origin of coordinates ($\eta_0=0$). In the simulations,
$w_0=5\lambda$ gives $\varepsilon=1/5\pi\ll1$. Hence, terms in the Lax series up
to ${\cal O}(\varepsilon^2)$ only need to be retained. The simulations demonstrate clearly
that the two models predict exit kinetic energies that agree to within 2$-$3\% for
the parameters used in Fig. \ref{fig2}. The linear scaling of the exit kinetic
energy with the laser intensity given by the plane-wave model only holds true for
intensities up to about $5\times10^{21}$ W/cm$^2$. For highly relativistic particles
the exit kinetic energy scales as the square root of the laser intensity,
which is the case in conventional TNSA \cite{plasma4}, too.

In Fig. \ref{fig2} (a) and (b) the exit kinetic energy $K$ of a single proton is displayed
as a function of $b$ in the limit of large $\eta$ and for optimal $b$ as a function of $\eta/2\pi$, respectively.
The energy gain is found to peak globally
at the chirp parameter values of $b\sim-0.003033$ and $-0.002979$ for the
plane-wave and focused pulses, respectively, and for the specific laser system
parameter set used. Other optimal chirp parameter values, which can be
determined in like fashion, lead to local, less pronounced kinetic energy
maxima. The fact that the optimal values of $b$ are so close (and so are the
corresponding exit kinetic energies) demonstrates that the analytic solution can
be quite reliable and that the plane-wave representation closely describes the
physics involved, at least for the parameter set used in which $w_0\gg\lambda$.
On the other hand, it is obvious that dependence upon $b$ of the
exit particle kinetic energy is sensitive, especially in the regime where high energy is
sought. In fact, the plane-wave calculation yields the uncertainty $\delta
K\approx1/2 ({\partial^2 K}/{\partial b^2})|_{b}(\delta b)^2$, which results from an
uncertainty $\delta b\sim10^{-6}$ in determinig $b$. The plane-wave
calculation yields $\delta K\sim0.195$ MeV, or about 0.074\%. When one takes $\delta b\sim10^{-5}$ instead, the uncertainties increase by two orders of magnitude ($\delta K\sim19.5$ MeV and 7.4\%).

Having studied the single-particle aspects, we now present
typical 2D3V-PIC (two spatial dimensions and three momentum space degrees of freedom)
simulation results of the interaction of the
chirped laser pulse with a pre-ionized hydrogen target.
The target might be either an expanding hydrogen cluster available at
sizes ranging from 1 nm to 1 $\mu$m \cite{silva2} or part of a hydrogen gas jet.
We are using the following simulation environment:
The spatial resolution of our simulation box is given by $\Delta z=\Delta x=
\lambda/100$, where the laser wavelength is still assumed to be $\lambda=1~\mu$m.
The particle number per cell is 100, for both protons and electrons.
The $x$-polarized laser enters the simulation box from the left and propagates in
the $z$-direction.
For the fields we choose a simple modification of the plane wave fields
given by $\bm{E}=\hat{\bm{x}}E_0f e^{-x^2/(2 w_0^2)}$ and
$\bm{B}=\hat{\bm{y}}(E_0/c)f e^{-x^2/(2 w_0^2)}$, with $f$ from Eq. (\ref{f})
and the factor $e^{-x^2/(2 w_0^2)}$ mimicking spatial focusing in the
polarization direction.
The target dimensions are assumed to have a length of
$0.2\lambda$ in the laser propagation direction, and extension $0.6\lambda$ in the
transverse direction.
Here, we assume the electrons
(in the hydrogen gas) to be underdense $n_e=0.1~n_c$, where
$n_c=1.1\times10^{21}$ cm$^{-3}$ is the critical density for $\lambda=1 \mu$m wavelength.
For the protons, the total number per shot being accelerated as one bunch amounts to $\approx10^7$.

\begin{figure}[t!]%
 \begin{minipage}{0.45\textwidth}
  \includegraphics[width=\textwidth]{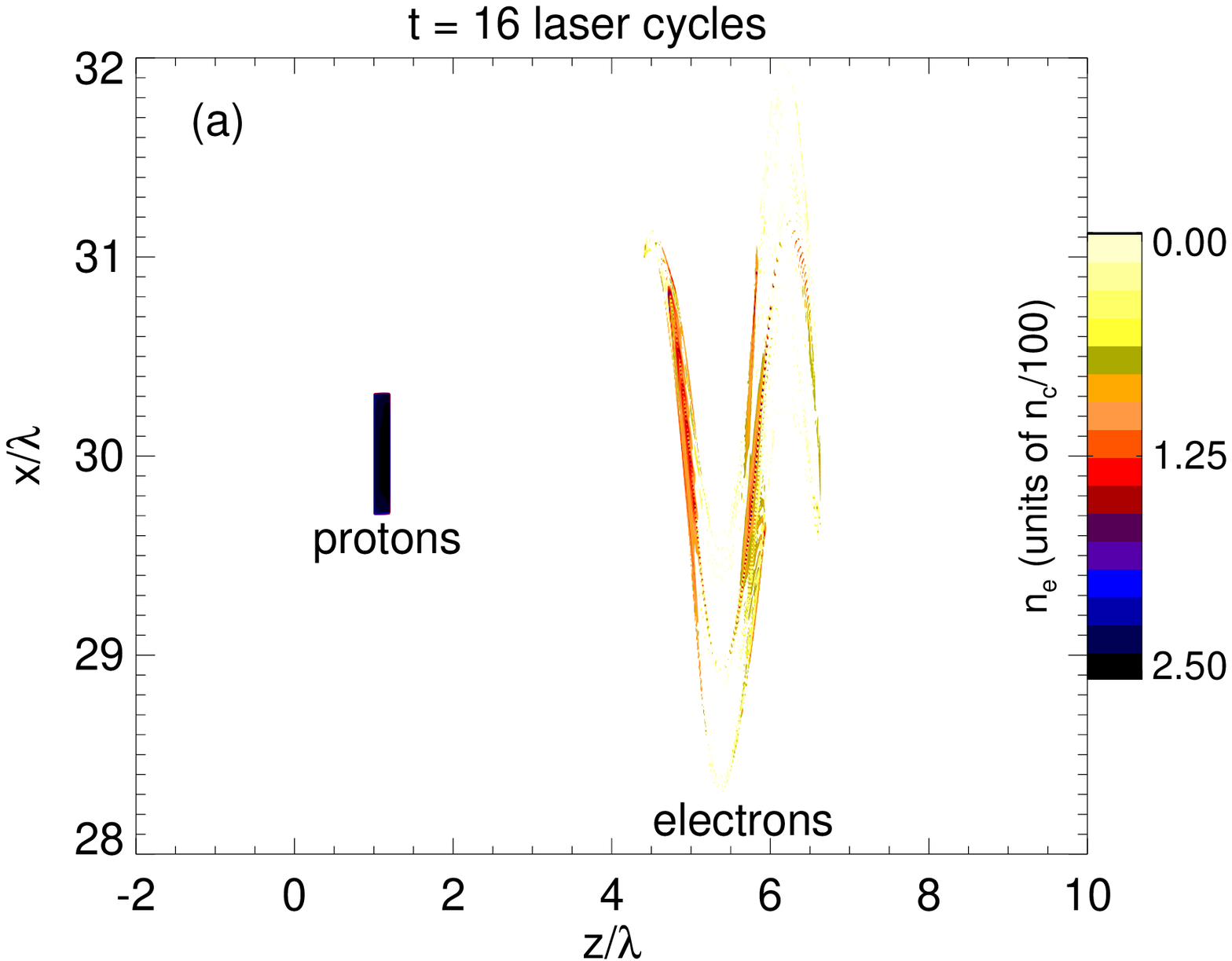}
 \end{minipage}
\begin{minipage}{0.45\textwidth}
 \includegraphics[width=\textwidth]{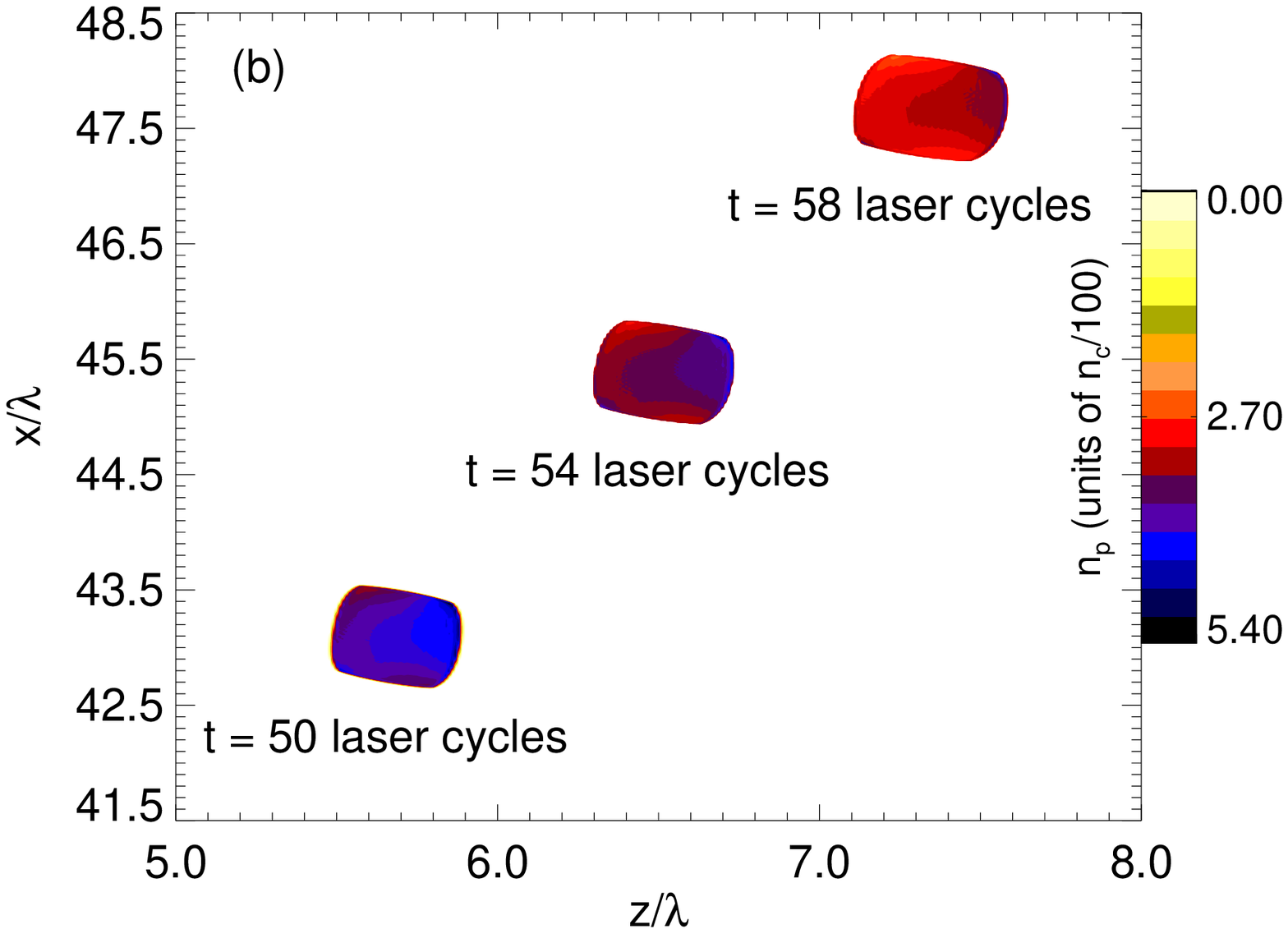}
 \end{minipage}

\caption{(color online). Snap-shots (a) of the electron and proton density distribution
during laser--plasma interaction and (b) of the proton density distribution after
laser--plasma interaction for various times.
The laser peak intensity is $2.55\times10^{21}$ W/cm$^2$ with
$w_0=5\lambda$, $\tau=50$ fs and $\lambda=1~\mu$m.}
   \label{fig3}%
\vspace{-2mm}
\end{figure}

Fig. \ref{fig3} provides an insight into the acceleration mechanism and the
plasma dynamics. From Fig. \ref{fig3} (a), we
can see that when the laser pulse approaches the gas target, the electrons get blown off the
target. The density profile of the electrons follows approximately the laser field
oscillation. At this time ($t = 16 $ laser cycles)
the proton distribution almost maintains its initial shape. After the
interaction of the gas target with the laser pulse,
we see from Fig. \ref{fig3} (b) that the protons are accelerated as a symmetric
bunch of homogeneous density (average density at $t = 50$ laser cycles: $\bar{n}_p\approx0.04~n_c$).
Note that the proton bunch is emitted at $70^\circ$ relative to the laser propagation direction ($z$-direction). 
Due to the comparatively low density of
the plasma target and, hence, suppressed Coulomb explosion,
the time-dependent beam divergence is low. For a macroscopic distance
of 30 cm, the cross-sectional area of the beam is about $0.98$ cm$^2$.
This number roughly corresponds to a divergence angle (opening angle of the cone defined
by the beam) of only $2^\circ$.
From the particle dynamics it is obvious that the prevailing part of the proton's kinetic energy is
transferred via direct interaction with the laser field.

\begin{figure}
\includegraphics[width=7.5cm]{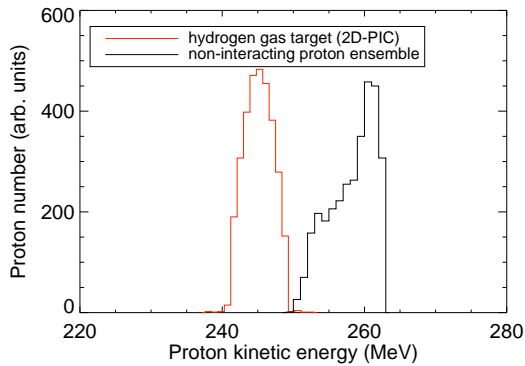}
\caption{(color online). Proton kinetic energy distributions after the interaction of a chirped laser pulse
($b=-0.003033$) with a hydrogen gas target (left peak) and with an
ensemble of protons without particle-particle interaction effects (right peak).
The laser peak intensity is $2.55\times10^{21}$ W/cm$^2$ for both cases.
}
\label{fig4}
\vspace{-2mm}
\end{figure}

In Fig. \ref{fig4} we compare the exit kinetic
energy distribution of 3000 initially randomly distributed non-interacting particles (cf. \cite{sal-prl2,galow} for the method)
in a spatial volume with the same dimensions as the gas target (cf. Fig \ref{fig3} (a)), with the one stemming from the laser --
gas target interaction.
In the case of pure vacuum acceleration
$K=258.3$ MeV $\pm~1.2\%$, while for the laser-plasma-cell acceleration
$K=245.2$ MeV $\pm~0.8\%$. The discrepancy between the reported mean kinetic energies is about $5\%$, which
can be attributed to particle-particle interaction effects. Moreover, the
simulation of the randomly distributed non-interacting particle ensemble has been
carried out in three spatial dimensions, whereas the PIC simulation is two dimensional.
Enlarging the initial plasma distribution leads
to an increase in the energy spread and to a decrease in the mean kinetic energy
of the created proton beam. For example a gas target with initial length of
$1\lambda$ (5 times larger than before) yields $K=245.1$ MeV $\pm~5.4\%$. However,
employing some velocity filter one could maintain the beam quality.
In order to reach similar particle energies by using an unchirped laser pulse one would have to raise the laser intensity by
three orders of magnitude, to $10^{24}$W/cm$^2$. At such an intensity the chirped
laser scheme (employing the same parameters as in Fig. \ref{fig2}) already yields monoenergetic proton beams with $K=17.3$ GeV $\pm~1.0\%$.

In summary, we have demonstrated the theoretical feasibility of creating a dense proton beam ($10^7$ protons
per bunch) of high energy ($\approx 250$ MeV) and good quality (energy spread $\sim1\%$),
from the interaction of a chirped laser pulse with an underdense hydrogen gas target.
The required laser peak intensity of about $10^{21}$ W/cm$^2$ is within the range
of state-of-the-art high-intensity laser systems \cite{bahk}.
Furthermore, we developed an analytical model for determining the optimal
pulse shape and the final kinetic energy of the particles, which is in good
agreement with the performed 2D-PIC simulations. Our work has mainly been 
concerned with proton acceleration to energies 
and densities required for hadron therapy. Following acceleration, ion  
beam shaping \cite{Schollmeier} beyond the scope of the present Letter has to be applied.

\begin{acknowledgments}
BJG acknowledges discussions with T.~Pfeifer.
YIS is supported by the Arab Fund for Economic and Social Development (State of
Kuwait). Supported by
Helmholtz Alliance HA216/EMMI.

\end{acknowledgments}

\end{document}